\documentclass[nofootinbib,notitlepage,superscriptaddress,10pt,aps,prd,twocolumn]{revtex4-1}

\usepackage{amsmath,mathtools}
\usepackage{tensor}
\usepackage{amssymb}
\usepackage{graphicx}
\usepackage{verbatim}
\newcommand{\leri}[1]{\left(#1\right)}
\newcommand{\lerisq}[1]{\left[#1\right]}

\begin{document}

\title{Big Bounce cosmology for Palatini $R^2$ gravity with a Nieh-Yan term}
\author{Flavio Bombacigno}
\email{flavio.bombacigno@uniroma1.it}
\affiliation{Physics Department, ``Sapienza'' University of Rome, Piazzale Aldo Moro 5, 00185 (Roma), Italy}
\author{Giovanni Montani}
\email{giovanni.montani@enea.it}
\affiliation{ENEA, FSN-FUSPHY-TSM, R.C. Frascati, Via E. Fermi 45, 00044 Frascati, Italy.\\
Physics Department, ``Sapienza'' University of Rome, P.le Aldo Moro 5, 00185 (Roma), Italy}

\begin{abstract}
We analyze the cosmological implementation of Palatini $f(R)$ theories, constructed with a Nieh-Yan term and solved with respect to the torsion. We consider the relevant case of the quadratic correction to the Hilbert-Palatini action in the Ricci scalar, mimicking the Starobinsky model of the metric formulation. We point out the emergence of peculiar cosmological scenarios, depending on the sign of such correction, able to reproduce bouncing settings and to restore the standard Universe dynamics in the late asymptotic limit. Furthermore, we outline the settling of Little-Rip dynamics, which calls for a deeper investigation in order to be regularized via matter creation. Finally, we also show that in our model the Immirzi field is asymptotically frozen in time, resembling the morphology of Loop Quantum Gravity standard formulation.
\end{abstract}

\maketitle

\section{Introduction}
General Relativity is a very rigorous and self-consistent construction for the geometrical representation of the gravitational interaction, from the point of view of the kinematic theory. In this sense, the tensor language arises as the mathematical implementation of the General Relativity Principle and the geodesic motion as the natural implication of the Equivalence Principle \cite{CQG}. 
\\ However, the Einsteinian dynamics, associated to the Einstein-Hilbert action is physically grounded only from the point of view of being the simplest choice, leading to equations which contain second derivatives of the metric tensor field. Indeed, simple generalizations of the gravitational action can be easily constructed by adding other scalar invariants to the Ricci scalar \cite{Nojiri:2010wj}, and of particular impact over the last two decades it has been the so-called $f(R)$ gravity, where the Ricci scalar is replaced by a generic function of it \cite{Sotiriou:2008rp, Nojiri:2017ncd}. Moreover, the success of this revised gravitational theory must be also 
attributed to the possibility of a Brans-Dicke reformulation \cite{ST} of the $f(R)$ theory in the so-called Jordan frame. 
\\ Besides the peculiar form of the action, another non-trivial ambiguity concerning the gravitational interaction is the possibility to consider \textit{a priori} the affine connection as an independent variable with respect to the metric field (Palatini or first order formulation) \cite{Pal}. Indeed, if the metric is related to the local causal structure of the space-time, the connection is in general a different geometric object, responsible for the transport of tensorial quantities across the space-time manifold. In general relativity, taking the affine connection as an independent entity leads to a dynamically equivalent description as it can completely solved in terms of the metric, \textit{i.e.} one simply recovers the Levi-Civita connection. However, when the Palatini formalism is implemented for a $f(R)$ model, although the connection could be still considered an auxiliary field devoid of a proper dynamics, its form is affected by the specific form of the function $f(\cdot)$ \cite{Sotiriou:2008rp}. Especially, it can be seen that the additional contribution due to the function $f(\cdot)$ allows to recast Palatini $f(R)$ theories into metric ones endowed with torsion \cite{Olmo:2011uz}. This means that the affine connection is equipped with an antisymmetric component depending of the function $f(\cdot)$, and we deal with a Riemann-Cartan space-time \cite{Cartan1,Cartan2,Hehl:2013qga, Hehl:1976kj,Shapiro:2001rz}.
\\ Now, since in Palatini $f(R)$ models torsion naturally emerges, in \cite{Bombacigno:2018tbo} we proposed the idea that for formulating $f(R)$ gravity in the connection language we have to include torsional contribution already into the Lagragian. In particular, also in relation with features of Loop Quantum Gravity (LQG) formalism 
\cite{Ashtekar1,Barbero1,Holst:1995pc,Immirzi:1996di,Thiemann:2007zz,Rovelli}, we considered in the Lagrangian 
density a Nieh Yan term \cite{NY1,NY2,Mercuri:2007ki} with the Immirzi parameter promoted to be a field \cite{Rovelli:1997na,Calcagni:2009xz,TorresGomez:2008fj,Cianfrani:2009sz,Bombacigno:2016siz}. 
\\ This choice allowed to fully solve torsion in terms of the function $f(\cdot)$ and the Immirzi field, reducing the original model to a scalar-tensor theory, characterized by an interesting phenomenology for the gravitational waves polarizations \cite{Bombacigno:2018tih}.
\\ Here, we explore the cosmological implementation of the theory proposed in \cite{Bombacigno:2018tbo} for a flat Friedman Universe, in order to shed light on the dynamical and physical implications that our revised Palatini $f(R)$ theory can have on the Universe history.
\\ Specifically, we analyze the $f(R)=R+\alpha R^2$ model, outlining very different evolutionary scenarios according the sign of the parameter $\alpha$, ruling the correction term to the Palatini-Hilibert part of the action \cite{Antoniadis:2018ywb,Tenkanen:2019jiq,Edery:2019txq}. \\For $\alpha>0$ we obtain a modified Friedman dynamics, marked by an effective gravitational constant depending on the matter content considered, and restoring the general relativity framework in the asymptotic limit. \\ Much more intriguing turns out to be the case $\alpha<0$, where we stress the settling of bouncing cosmologies as a purely classic effect, due to the non minimal coupling between the Immirzi field and the extended gravity sector of the theory. In this case our analysis also points out the existence of closed Universe solutions, even in the absence of spatial curvature, still affected by the singularity and where general relativity is never restored. Eventually, it is worth noting that in our model the Immirzi field can be dynamically relaxed to a constant by the Universe expansion, recovering the standard LQG perspective.
\\ The paper is structured as follows. In Sec.~\ref{sec2} we discuss Palatini $f(R)$ gravity and the role played by torsion in this framework; in Sec.~\ref{sec3} we briefly recall the main features of our model in the general case, and in Sec.~\ref{sec4} we specialize to the isotropic and homogeneous background. In Sec.~\ref{sec5} we consider the implications of the correction term in the Universe evolution. Finally, in Sec.~\ref{sec6} conclusions are drawn.
\section{The role of torsion in Palatini  $f(R)$ theories}\label{sec2}
The action for generic $f(R)$ models in Palatini formulation is given by\footnote{We set $\chi=8\pi G$ and $c=1$.}
\begin{equation}
S=\frac{1}{2\chi}\int d^4x\;\sqrt{-g}\,f(R)+S_M[g_{\mu\nu},\psi],
\label{f(R)action}
\end{equation}
where $S_M$ represents the matter action and $\psi$ collects globally the matter fields, which minimally couples to the metric field only. The function $f(R)$ depends on the Ricci scalar $R$, which according a first order analysis reads as
\begin{equation}
R=g^{\mu\nu}R_{\mu\nu}(\Gamma,\partial\Gamma)=g^{\mu\nu}R\indices{^{\rho}_{\mu\rho\nu}}(\Gamma,\partial\Gamma),
\label{RicciScalar}
\end{equation}
the Riemann tensor $R\indices{^{\mu}_{\nu\rho\sigma}}$ being function of the affine connection and its derivative, \textit{i.e.}
\begin{equation}
R\indices{^{\mu}_{\nu\rho\sigma}}=\partial_{\rho}\Gamma\indices{^\mu_{\nu\sigma}}-\partial_{\rho}\Gamma\indices{^\mu_{\nu\rho}}+\Gamma\indices{^\mu_{\tau\rho}}\Gamma\indices{^\tau_{\nu\rho}}-\Gamma\indices{^\mu_{\tau\sigma}}\Gamma\indices{^\tau_{\nu\rho}}.
\end{equation}
It is worth remarking that the form of the connection is not established in the well-know Levi-Civita solution, as in the metric approach, but is determined properly by the equation of motion obtained from \eqref{f(R)action}. Indeed, if we assume the affine connection to be symmetric in its lower indices, which \textit{a priori} could be not guaranteed, the variation of \eqref{f(R)action} with respect to the metric field leads to
\begin{equation}
f'(R)R_{(\mu\nu)}-\frac{1}{2}f(R)=\chi T_{\mu\nu},
\label{f(R)_g_eq}
\end{equation}
with a prime denoting differentiation with respect to the argument and brackets symmetrization on the indices. The equation for the connection is given instead by
\begin{equation}
\nabla_{\rho}\leri{\sqrt{-g}f'(R)g^{\mu\nu}}=0,
\label{connectP2}
\end{equation}
where $\nabla_{\mu}$ is the covariant derivative from $\Gamma\indices{^\rho_{\mu\nu}}$ \textbf{and} the stress-energy tensor $T_{\mu\nu}$ is defined as
\begin{equation}
T_{\mu\nu}\equiv -\frac{2}{\sqrt{-g}}\frac{\delta S_M}{\delta g^{\mu\nu}}.
\end{equation}
Eventually, condition \eqref{connectP2} can be still restated as the Levi-Civita definition for the connection $\Gamma\indices{^\rho_{\mu\nu}}$, provided we perform a conformal transformation of the metric $g_{\mu\nu}$, that is
\begin{equation}
\tilde{g}_{\mu\nu}\equiv f'(R)g_{\mu\nu}.
\end{equation}
Then, a solution for \eqref{connectP2} is given by
\begin{equation}
\begin{split}
\Gamma\indices{^\rho_{\mu\nu}}=&\frac{1}{2}\tilde{g}^{\rho\sigma}\leri{\partial_{\nu}\tilde{g}_{\mu\sigma}
+\partial_{\mu}\tilde{g}_{\nu\sigma}-\partial_{\sigma}\tilde{g}_{\mu\nu}}=\\
=&\frac{1}{2}g^{\rho\sigma}\leri{\partial_{\nu}g_{\mu\sigma}
+\partial_{\mu}g_{\nu\sigma}-\partial_{\sigma}g_{\mu\nu}}+\\
+&\frac{1}{2}\leri{\delta\indices{^\rho_\mu}\partial_{\nu}\ln f'(R)
+\delta\indices{^\rho_\nu}\partial_{\mu}\ln f'(R)-g_{\mu\nu}\partial^\rho\ln f'(R)}
.\label{LeviCivita_pal}
\end{split}
\end{equation}
However, if we do not impose at the very beginning any particular symmetries properties on the form on the connection, solution \eqref{LeviCivita_pal} is not the most general form the connection we can have \cite{Olmo:2011uz}. In fact, before variation of the action be performed, we expect that connection could be endowed with an antisymmetric component,  \textit{i.e.} torsion tensor might be present:
\begin{equation}
T\indices{^\rho_{\mu\nu}}\equiv\frac{1}{2}\leri{\Gamma\indices{^\rho_{\mu\nu}}-\Gamma\indices{^\rho_{\nu\mu}}}\neq 0,
\end{equation}
and explicit calculations show that the solution \eqref{LeviCivita_pal} has to be enlarged to include a vector-like contribution, namely
\begin{equation}
\Gamma\indices{^\rho_{\mu\nu}}\rightarrow\Gamma\indices{^\rho_{\mu\nu}}-\frac{2}{3}\delta\indices{^\rho_\mu}V_\nu.
\label{vector connection}
\end{equation}
Now, since the symmetric part of the Ricci tensor $R_{(\mu\nu)}$ is not affected by \eqref{vector connection}, neglecting or considering any torsional contribution due to $V_\nu$ seems to be dynamically equivalent and we always recover the metric field equation \eqref{f(R)_g_eq}. However, by the inspection of \eqref{LeviCivita_pal} it is clear that also the function $f(R)$ is responsible for the vector part
\begin{equation}
-\frac{2}{3}\delta\indices{^\rho_\mu}V^{f(R)}_\nu=\frac{1}{2}\delta\indices{^\rho_\mu}\partial_\nu\ln f'(R),
\end{equation}
so we can imagine to fix $V_\nu=-V^{f(R)}_\nu$, in order to deal with a total vanishing vector component in the connection. This choice allows us to recast \eqref{LeviCivita_pal} in the more suitable form
\begin{equation}
\Gamma\indices{^\rho_{\mu\nu}}=\bar{\Gamma}\indices{^\rho_{\mu\nu}}+K\indices{^\rho_{\mu\nu}},
\label{connection_contorsion}
\end{equation}
where $\bar{\Gamma}\indices{^\rho_{\mu\nu}}$ is the ordinary Levi-Civita connection, defined in terms of the metric field $g_{\mu\nu}$, and $K\indices{^\rho_{\mu\nu}}$ the so-called contorsion tensor related to torsion by:
\begin{equation}
K\indices{^\rho_{\mu\nu}}=\frac{1}{2}\leri{T\indices{^\rho_{\mu\nu}}-T\indices{_\mu^\rho_\nu}-T\indices{_\nu^\rho_\mu}},
\end{equation}
that from \eqref{LeviCivita_pal} is recognized in
\begin{equation}
K_{\rho\mu\nu}=\frac{1}{2}\leri{g_{\rho\nu}\partial_\mu\ln f'(R)-g_{\mu\nu}\partial_\rho\ln f'(R)}.
\label{f(R) contorsion}
\end{equation}
Then, since the contorsion in general still depends on $f'(R)$, the definition \eqref{f(R) contorsion} (or \eqref{LeviCivita_pal} itself), is well-posed only if one is able to express $R$ as a function of quantities which do not depend on the connection. With this regard, if we trace the equation for the metric field \eqref{f(R)_g_eq}, we obtain the structural equation:
\begin{equation}
f'(R)R-2f(R)=\chi T,
\label{f(R)pal_trace}
\end{equation}
which in principle can be solve algebraically for $R=R(T)$, allowing us to completely determine the connection in terms of the metric field and the matter source. Now, taking into account \eqref{connection_contorsion}, the equation for the gravitational field can be rearranged as:
\begin{equation}
\begin{split}
\bar{G}_{\mu\nu}(g)=&\frac{\chi}{f'(R)}T_{\mu\nu}-\frac{1}{2}g_{\mu\nu}\leri{R-\frac{f(R)}{f'(R)}}+\\
-&\frac{3}{2f'(R)^2}\leri{\bar{\nabla}_\mu f'(R)\bar{\nabla}_\nu f'(R)-\frac{1}{2}g_{\mu\nu}(\bar{\nabla} f'(R))^2}+\\
+&\frac{1}{f'(R)}g_{\mu\nu}\leri{\bar{\nabla}_\mu\bar{\nabla}_\nu-g_{\mu\nu}\bar{\Box}}f'(R),
\end{split}
\label{f(R) effective g eq}
\end{equation}
where $\bar{G}_{\mu\nu}$ is the Einstein tensor constructed via the metric field $g_{\mu\nu}$ only, the d'Alambert operator defined by $\bar{\Box}\equiv g^{\mu\nu}\bar{\nabla}_\mu\bar{\nabla}_\nu$, and $\leri{\bar{\nabla}f'(R)}^2\equiv\bar{\nabla}_\mu f'(R)\bar{\nabla}^\mu f'(R)$.
\\ In \eqref{f(R) effective g eq}, being ultimately $f'(R)$ a function of $T$, derivatives of the stress-energy tensor appear, changing how the matter source generates space-time curvature. Moreover, when $T=0$, as in vacuum, relation \eqref{f(R)pal_trace} may admit a set of constant solutions $R=R_0^{(i)}$ and in this case equation \eqref{f(R) effective g eq} simply reduces to GR equation with an effective cosmological constant
\begin{equation}
\Lambda_0=\frac{1}{2}\leri{R_0-\frac{f(R_0)}{f'(R_0)}}.
\end{equation}
Therefore, the scenarios offered by Palatini formulation of $f(R)$ theories depart significantly from standard predictions just in the presence of matter, where the connection is not an independent variable, but an auxiliary field affecting the way metric and matter interact. Furthermore, by virtue of \eqref{connection_contorsion}, we see that first order $f(R)$ models can equivalently restated as metric theories endowed with torsion, primarily given by the specific form of the function $f$. However, if torsion is present at very fundamental level, it seems reasonable to include torsional contributions already into the Lagrangian. In this respect, a simple way for achieving that in a LQG-oriented analysis is offered by $f(R)$ extensions of the well-known Nieh-Yan action, which represents the starting point of our work.

\section{The Nieh-Yan $f(R)$ model}\label{sec3}
Let us consider the following extension of the action \eqref{f(R)action} (see \cite{Bombacigno:2018tbo}), where the so-called Nieh-Yan term is considered in the presence of a dynamical Immirzi field $\beta(x)$:
\begin{equation}
\begin{split}
S_{NY}=&\frac{1}{2\chi}\int d^4x\;\sqrt{-g}\,f(R)+\\
+&\frac{1}{4\chi}\int d^4x\;\sqrt{-g}\,\beta(x)\epsilon^{\mu\nu\rho\sigma}\leri{g_{\tau\lambda}T\indices{^{\tau}_{\mu\nu}}T\indices{^{\lambda}_{\rho\sigma}}-R_{\mu\nu\rho\sigma}}+\\
+&S_M[g_{\mu\nu,} \psi].
\label{NY_f(R)_action}
\end{split}
\end{equation}
When $f(R)=R$, action \eqref{NY_f(R)_action} resembles for $\beta$ constant the Nieh-Yan action usually adopted  in Loop Quantum Gravity, and the Immirzi parameter rules a total divergence that does not affect classically the equations of motion. However, as stressed in \cite{Calcagni:2009xz,Bombacigno:2016siz,Bombacigno:2018tih}, if $\beta$ is space-time dependent, it behaves as a source of torsion and the theory can be reformulated as General Relativity in the presence of a minimally coupled massless scalar field. Therefore, if we keep $f(R)\neq R$ generic, we expect that both types of torsion could generate a dynamic larger framework with respect to \eqref{f(R) effective g eq}, by virtue of the non trivial coupling between the Immirzi field and the gravitational degrees of freedom (d.o.f).
\\ Then, following the analysis of \cite{Bombacigno:2018tbo} that here we are widening for including matter as well, with a bit of algebra action \eqref{NY_f(R)_action} can be rewritten in the more convenient scalar-tensor form:
\begin{equation}
\begin{split}
S_{NY}=&\frac{1}{2\chi}\int d^4x\;\sqrt{-g}\leri{\phi \bar{R}-g^{\mu\nu}\Pi_{\mu\nu}(\phi,\beta)-V(\phi)}\\&+S_M[g_{\mu\nu,} \psi],
\label{scal_tens_action}
\end{split}
\end{equation}
where $\phi\equiv f'(R)$ and $\bar{R}$ represents the metric Ricci scalar, depending only on metric variables. In particular, we introduced the quantities
\begin{align}
&\Pi_{\mu\nu}(\phi,\beta)=\frac{3}{2\phi}\leri{\bar{\nabla}_{\mu}\beta\bar{\nabla}_{\nu}\beta-\bar{\nabla}_{\mu}\phi\bar{\nabla}_{\nu}\phi}
\label{Def Pi}\\
&V(\phi)\equiv \phi R(\phi)-f(R(\phi)).
\label{Potential}
\end{align}
By analogy with \eqref{f(R) contorsion}, dealing at the effective level with \eqref{scal_tens_action} means having a contorsion tensor given by
\begin{equation}
K_{\rho\mu\nu}=\frac{1}{2\phi}\leri{g_{\rho \nu}\bar{\nabla}_{\mu}\phi-g_{\mu \nu}\bar{\nabla}_{\rho}\phi}-\frac{1}{2\phi}\epsilon_{\rho\mu\nu\sigma}\bar{\nabla}^{\sigma}\beta,
\label{NY_f(R)_contorsion_sol}
\end{equation}
which for $\beta$ constant boils down to \eqref{f(R) contorsion}.
\\ Then, varying \eqref{scal_tens_action} with respect to $g_{\mu\nu}$ carries out:
\begin{equation}
\begin{split}
\bar{G}_{\mu\nu}=&\frac{\chi}{\phi}T_{\mu\nu}-\frac{1}{2\phi}g_{\mu\nu}V(\phi)+\\+&\frac{1}{\phi}\leri{\Pi_{\mu\nu}(\phi,\beta)-\frac{1}{2}g_{\mu\nu}\Pi\indices{^\rho_\rho}(\phi,\beta)}+\\
+&\frac{1}{\phi}\leri{\bar{\nabla}_{\mu}\bar{\nabla}_{\nu}-g_{\mu\nu}\bar{\Box}}\phi
,
\label{eq_g}
\end{split}
\end{equation}
while the equations for $\phi$ and $\beta$ are given by, respectively:
\begin{equation}
\bar{R}=-\frac{3}{2\phi^2}\leri{\bar{\nabla}_{\mu}\beta\bar{\nabla}^{\mu}\beta+\bar{\nabla}_{\mu}\phi\bar{\nabla}^{\mu}\phi}+\frac{3\bar{\Box}\phi}{\phi}+V'(\phi)
\label{eq_phi}
\end{equation}
and
\begin{equation}
\bar{\Box}\beta(x)=\frac{\bar{\nabla}_{\mu}\beta(x)\bar{\nabla}^{\mu}\phi}{\phi}.
\label{eq_beta}
\end{equation}
Substituting in \eqref{eq_phi} the trace of \eqref{eq_g}, we get the modified structural equation (see \eqref{f(R)pal_trace} for comparison):
\begin{equation}
2\; V(\phi)-\phi \;V'(\phi)=\chi T-\frac{3\bar{\nabla}_\mu\beta\bar{\nabla}^\mu\beta}{\phi}.
\label{tracephi}
\end{equation}
that once we chose the specific $f(R)$ model, allows us to solve for $\phi=\phi((\bar{\nabla}\beta)^2,T)$. We point out that still in vacuum relation \eqref{tracephi} admits a larger set of solutions for $\phi$, which is not compelled to relax to a constant value as in original formulation. The reason for this is the non trivial coupling between the Immirzi field and the enlarged gravitational sector, which makes the vacuum configuration never actually devoid of matter content. In particular, the Immirzi term in \eqref{tracephi} only partially resembles the contribution to the structural equation of scalar fields in ordinary Palatini $f(R)$ gravity, where we would expect a standard kinetic term deprived of the coupling with $\phi$. Furthermore, we note that the requirement of recovering to some extent a proper vacuum state, as it is described by \eqref{f(R)pal_trace} for $T=0$, raises the issue about the relaxation of the Immirzi field on a constant configuration, able to reproduce standard LQG theory as well. In this respect, these problems can be properly addressed in cosmology, where we can ask for the asymptotically freezing of the Immirzi field during the expansion of the Universe. Of course, such a mechanism does not accounts for local fluctuations $\delta\beta(x)$ (see \cite{Bombacigno:2018tbo,Bombacigno:2018tih}), but it can be considered responsible for the evolution of the background value $\beta_B$, where $\beta=\beta_B+\delta\beta$, which we may demand to match with independent LQG estimates \cite{Ghosh:2004wq,Ghosh:2011fc}.

\section{Modified Friedmann equation}\label{sec4}
A simple model for analyzing the role played by $\beta_B$ is represented by the homogeneous and isotropic Universe, described by the Friedman-Robertson-Walker (FRW) line element\footnote{We set the lapse function $N(t)=1$.}
\begin{equation}
ds^2=-dt^2+a^2(t)\left(\frac{dr^2}{1-kr^2}+r^2(d\theta^2+\sin^2\theta d\phi^2)\right),
\label{FRWmetric}
\end{equation}
the scale factor $a(t)$ being the only metric dynamical degree of freedom and $k$  the curvature of space. Within such a framework, the background value for the Immirzi field can be considered function of the cosmological time $t$ only, namely $\beta_B=\beta_B(t)$. Now, be $T_{\mu\nu}$ the stress-energy tensor for a perfect fluid, \textit{i.e.}:
\begin{equation}
T_{\mu\nu}=\leri{\rho+P}u_\mu u_\nu+g_{\mu\nu}P,
\label{str_en_tens}
\end{equation}
where $\rho$ and $P$ are the energy density and the pressure, respectively, and $u_\mu=(-1,0,0,0)$. Then, in the presence of the energy density $\rho$ the Friedman equation stemming from \eqref{eq_g} can be rearranged as
\begin{equation}
H^2=\leri{\frac{\dot{a}}{a}}^2=\frac{\chi}{3\phi}\leri{\rho_{eff}+\rho}-\frac{k}{a^2},
\label{CE1}
\end{equation}
with dot denoting time derivative, while combining the equation for the $ii$ component with \eqref{CE1} we get the acceleration equation
\begin{equation}
\frac{\ddot{a}}{a}=-\frac{\chi}{6\phi}\leri{\rho+\rho_{eff}+3(P+P_{eff})},
\label{CE2}
\end{equation}
where we introduced the effective energy density and pressure given by, respectively:
\begin{equation}
\rho_{eff}\equiv\frac{1}{\chi}\lerisq{\frac{3}{4\phi}\leri{\dot{\beta}_B^2-\dot{\phi}^2}+\frac{1}{2}V(\phi)-3\frac{\dot{a}}{a}\dot{\phi}}
\label{CE3}
\end{equation}
and
\begin{equation}
P_{eff}\equiv\frac{1}{\chi}\lerisq{\frac{3}{4\phi}\leri{\dot{\beta}_B^2-\dot{\phi}^2}-\frac{1}{2}V(\phi)+\ddot{\phi}+2\frac{\dot{a}}{a}\dot{\phi}}.
\label{CE4}
\end{equation}
Deriving equation \eqref{CE1} with respect to time and plugging \eqref{CE2} in it, we can obtain the relation
\begin{equation}
\begin{split}
&\dot{\rho}+3\leri{\frac{\dot{a}}{a}}\leri{\rho+P}+\\
&+\dot{\rho}_{eff}-\frac{\dot{\phi}}{\phi}\leri{\rho_{eff}+\rho}+3\leri{\frac{\dot{a}}{a}}\leri{\rho_{eff}+P_{eff}}=0.
\end{split}
\end{equation}
Then, in order the standard continuity equation be preserved, that is:
\begin{equation}
\dot{\rho}+3\frac{\dot{a}}{a}\leri{\rho+P}=0,
\label{CE5}
\end{equation}
the following condition has to be fulfilled
\begin{equation}
\dot{\rho}_{eff}-\frac{\dot{\phi}}{\phi}\leri{\rho+\rho_{eff}}+3\frac{\dot{a}}{a}\leri{\rho_{eff}+P_{eff}}=0.
\label{CE6}
\end{equation}
By virtue of \eqref{CE3} and \eqref{CE4} this relation can be rewritten as
\begin{equation}
\frac{\dot{\phi}}{2}\leri{V'(\phi)-\bar{R}+\frac{3\dot{\phi}^2}{2\phi}-\frac{3\dot{\beta}_B^2}{2\phi^2}+\frac{3\bar{\Box}\phi}{\phi}}-\frac{3\dot{\beta}_B\bar{\Box}\beta_B}{2\phi}=0,
\label{CE8}
\end{equation}
where we used the expressions of the Ricci scalar and the d'Alambert operator for the background \eqref{FRWmetric}. Hence, using \eqref{eq_beta}, relation \eqref{CE8} takes the form
\begin{equation}
\frac{\dot{\phi}}{2}\leri{V'(\phi)-\bar{R}+\frac{3\dot{\phi}^2}{2\phi}+\frac{3\dot{\beta}_B^2}{2\phi^2}+\frac{3\bar{\Box}\phi}{\phi}}=0,
\label{CE9}
\end{equation}
which is identically satisfied given \eqref{eq_phi}.
\\Therefore, if the equation of state $P=w\rho$ holds, where $w$ is the polytropic index, from \eqref{CE5} the standard solution can be obtained, namely
\begin{equation}
\rho(a)=\frac{\mu^2}{a^{3(w+1)}},
\label{CE7}
\end{equation}
with $\mu$ a constant. 
\\ Furthermore, we note that the equation \eqref{eq_beta} for $\beta_B$ can be actually solved analitically for $\dot{\beta}_B$. Indeed, for a FRW background, \eqref{eq_beta} simply reads as:
\begin{equation}
\ddot{\beta}_B(t)+\leri{3\frac{\dot{a}}{a}-\frac{\dot{\phi}}{\phi}}\dot{\beta}_B=0,
\end{equation}
whose solution is given by:
\begin{equation}
\dot{\beta}_B(t)=C_0\frac{\phi(t)}{a^3(t)},
\label{eq_beta_FRW}
\end{equation}
where we defined the integration constant $C_0\equiv\frac{\dot{\beta}_B(t_0)a(t_0)^3}{\phi(t_0)}$ for a fiducial instant $t_0$.
Thus, inserting \eqref{CE7} and \eqref{eq_beta_FRW} in \eqref{tracephi} yields:
\begin{equation}
2 V(\phi)-\phi V'(\phi)=\chi\mu^2\frac{3w-1}{a^{3(w+1)}}+\frac{3C_0^2}{a^6}\phi,
\label{T3}
\end{equation}
that, once a peculiar $f(R)$ model has been fixed, allows us to express $\phi$ as a function of the scale factor $a(t)$, by virtue of \eqref{Potential}. Moreover, given \eqref{eq_beta_FRW}, this implies in turn that $\dot{\beta}_B(t)$ depends on time only by means of the scale factor. Therefore, the requirement that the Immirzi field relaxes on a constant, can be equivalently restated as
\begin{equation}
\lim_{a\rightarrow +\infty}\dot{\beta}_B=0.
\label{relaxation immirzi}
\end{equation} 
Now, taking into account \eqref{eq_beta_FRW}, the Friedman equation can be reformulated as:
\begin{equation}
\begin{split}
H^2=&\frac{\chi}{3\phi}\frac{\mu^2}{a^{3(w+1)}}+\frac{C_0^2}{4a^6}+\frac{V(\phi)}{6\,\phi}-\frac{1}{4}\frac{\dot{\phi}^2}{\phi^2}-\frac{\dot{a}}{a}\frac{\dot{\phi}}{\phi}-\frac{k}{a^2}.
\end{split}
\label{fried_mod1}
\end{equation}
We note that, since now $\phi$ has to be understood by means of \eqref{T3} as a function of the scale factor $a$, the terms in the R.H.S. of \eqref{fried_mod1} depending on the time derivative of $\phi$ always give rise to terms proportional to $H^2$, regardless the $f(R)$ model considered. Therefore, it is possible to rearrange \eqref{fried_mod1} in the following way:
\begin{equation}
H^2=\frac{\leri{\frac{\chi\rho(a)}{3\phi}+\frac{C_0^2}{4a^6}+\frac{V(\phi)}{6\,\phi}-\frac{k}{a^2}}_{\phi=\phi(a)}}{F_1(a)^2},
\label{fried_mod2}
\end{equation}
where $F_1(a)$ is a function of the scale factor that has the general form
\begin{equation}
F_1(a)=\leri{1+\frac{a}{2}\frac{d \ln\phi(a)}{da}},
\end{equation}
and the term depending on the time derivative of the Immirzi field appears, by virtue of \eqref{eq_beta_FRW}, as a sort of scalar field energy density.\\ We note that the behaviour of $F_1(a)$ and $\phi(a)$ can remarkably affect the evolution of the scale factor. Indeed, in the presence of any polos and zeros for $F_1(a)$, the function $H$ can vanish or diverge, giving rise to peculiar cosmological scenarios. Similarly, by virtue of the coupling with the energy density content and the potential term, also $\phi(a)$ can be in principle responsible for analogous frameworks.

\section{The model $f(R)=R+\alpha R^2$}
\label{sec5}
In the following, we will restrict our attention on a specific Lagrangian, including a correction term  quadratic in the total Ricci scalar $R$, \textit{i.e.}:
\begin{equation}
f(R)=R+\alpha R^2.
\label{Staro_model}
\end{equation} 
It is worth noting that with respect to the metric approach (the well-established Starobinsky model \cite{Starobinsky,Hawking:2000bb,Kehagias:2013mya,Whitt:1984pd}), in Palatini formulation there are no issues concerning possible instabilities of the solution \cite{Olmo:2011uz,Sotiriou:2006sf}, being that ultimately due to the non dynamical nature of the field $\phi$. For this reason, the real parameter $\alpha$ is not compelled \textit{a priori} to be positive, and also negative values represent a suitable choice.
\\ Then, when the model \eqref{Staro_model} is taken into account, the potential $V(\phi)$ can be easily found, that is
\begin{equation}
V(\phi)=\frac{1}{\alpha}\leri{\frac{\phi-1}{2}}^2,
\label{Staro_potential}
\end{equation}
which inserted in \eqref{T3} gives us:
\begin{equation}
\phi=\frac{a^6\,f(a)}{a^6+6\alpha C^2_0},
\label{phi_a_R2}
\end{equation}
being $f(a)$ a function of the energy density, \textit{i.e.}
\begin{equation}
f(a)=1-2\alpha \chi (3w-1)\rho(a).
\end{equation} 
\\ Eventually, setting $k=0$, by means of \eqref{Staro_potential}--\eqref{phi_a_R2} the Friedman equation can be rearrange as
\begin{equation}
H^2=\frac{(a^6+6\alpha C_0^2)\leri{4\chi\rho+\frac{3C_0^2 f(a)}{a^6+6\alpha C_0^2}+\frac{2\alpha\leri{\chi(3w-1)a^6\rho+3C_0^2}^2}{(a^6+6\alpha C_0^2)^2}}}{12 a^6 f(a) \leri{\frac{a^6+24\alpha C_0^2}{a^6+6\alpha C_0^2}+\frac{af'(a)}{2f(a)}}^2}.
\label{modified friedman general}
\end{equation}
By first inspection of \eqref{modified friedman general}, we see that according the sign of $\alpha$ the parameter $C_0^2$, related to the Immirzi energy density, is crucial in determining the critical points of the Friedman equation. In particular, with the aim of investigating the possible emergence of bouncing cosmologies ruled by the Immirzi field \cite{Olmo:2008nf,Barragan:2009sq,Koivisto:2010jj}, it can be instructive to consider the vacuum case $\rho=0$, where $f(a)=1$ and \eqref{modified friedman general} takes a very simple form. More complex examples, even if still feasible for analytic studies, are represented both by the cosmological constant case, where $\rho$ is constant and the term $f'(a)$ in \eqref{modified friedman general} vanishes, and by the radiation one, where the trace of $T_{\mu\nu}$ is zero and $f(a)=1$ as in vacuum.
\subsection{The vacuum case}\label{vacuum}
The vacuum model constitutes a very useful tool for studying the effects, on the space-time structure, of the Immirzi coupling to gravitational d.o.f. . In this case relation \eqref{phi_a_R2} is simply
\begin{equation}
\phi(a)=\frac{a^6}{a^6+6\alpha C_0^2}
\label{phi a R2 vacuum}
\end{equation}
and \eqref{modified friedman general} boils down to
\begin{equation}
H^2=\frac{C_0^2}{4a^6}\frac{(a^6+6\alpha C_0^2)(a^6+12\alpha C_0^2)}{(a^6+24 \alpha C_0^2)^2}.
\label{friedman vacuum R2}
\end{equation}
When $\alpha>0$, equation \eqref{phi a R2 vacuum} does not exhibit critical points and it can be recast into the form
\begin{equation}
H^2=\frac{\chi_\beta(a)}{3}\rho_{\beta},
\label{R2_GR}
\end{equation}
which represents the Friedman equation for the scalar field energy density $\rho_{\beta}\equiv \frac{3 C_0^2}{4\chi a^6}$. It is characterized by an effective gravitational constant
\begin{equation}
\chi_\beta(a)\equiv \frac{(a^6+6\alpha C_0^2)(a^6+12\alpha C_0^2)}{(a^6+24 \alpha C_0^2)^2} \,\chi,
\label{G eff imm}
\end{equation}
and General Relativity is recovered for $a\rightarrow + \infty$, where $\chi_\beta\rightarrow \chi$ and $\phi\rightarrow 1$ in agreement\footnote{The value $\phi=1$ corresponds to $f'(R)=1$.} with \eqref{phi a R2 vacuum}.
\\ When instead $\alpha<0$, the presence of $C_0^2$ affects drastically \eqref{phi a R2 vacuum} . Indeed, in order the condition $H^2\ge 0$ be preserved, the scale factor $a$ cannot assume arbitrary values in $\mathbb{R}^+$, but is constrained into domains
\begin{equation}
\begin{split}
&\mathcal{D}_1:\;a \in [0,(-6\alpha C_0^2)^{1/6}],\\
&\mathcal{D}_2:\;a \in [(-12\alpha C_0^2)^{1/6},+\infty).
\label{vacuum domains}
\end{split}
\end{equation}
Therefore, we deal with two disconnected branches, denoting two different kind of Universe. In particular, the region $\mathcal{D}_1$ defines a closed Universe, even for $k=0$, bounded by the turning point $a_T=(-6\alpha C_0^2)^{1/6}$, where the General Relativity limit is never reached ($\phi=0$ for $a=0$ and $\phi\rightarrow -\infty$ for $ a\rightarrow a_T$) and it can be disregarded since unphysical. \\ Instead, the brach $\mathcal{D}_2$ is endowed with the critical point $a_B=(-12\alpha C_0^2)^{1/6}$ where $H^2=0$ and a bounce occurs, driven by the Immirzi energy density. That can be further proved by evaluating \eqref{CE2} at the bounce, where it can be recast into the form
\begin{equation}
\frac{\ddot{a}}{a}=\frac{-\frac{\chi (1+3w)}{6\phi}\rho-\frac{C_0^2}{2a^6}+\frac{V(\phi)}{\phi}-\frac{1}{2}H^2 F_2(a)}{F_1(a)},
\label{acceleration modified general}
\end{equation}
with $F_2(a)$ given by
\begin{equation}
F_2(a)\equiv a^2 \frac{d^2}{da^2}\ln \phi+a \frac{d}{da}\ln \phi.
\end{equation}
Now, since $\phi$ is not singular at $a=a_B$, at the bounce ($H^2=0$) relation \eqref{acceleration modified general} simply gives
\begin{equation}
\left.\frac{\ddot{a}}{a}\right |_{a=a_B}=-\frac{1}{32\alpha},
\end{equation}
which is positive for $\alpha<0$.
\\Remarkably, in this case \eqref{R2_GR} can be put in the LQC-like form
\begin{equation}
H^2=\frac{\chi}{3}\rho_\beta\leri{1-\frac{\rho_\beta}{\rho_{crit}^{vac}}},
\label{bounce vacuum}
\end{equation}
where we introduced the critical density
\begin{equation}
\rho_{crit}^{vac}\equiv\frac{(a^6+24\alpha C_0^2)^2}{8\alpha\chi a^6(5a^6+84\alpha C_0^2)}.
\end{equation}
With respect to \cite{Olmo:2008nf,Barragan:2009sq}, where analogous results were discussed, we stress that in our case we are able to reproduce bouncing cosmology for \eqref{Staro_model} also in the presence of stiff-like matter ($w=1$) (properly mimicked by the Immirzi field contribute) when $\alpha<0$. Moreover, requiring that the bounce occurs for Planckian energy density, allows to set the order of magnitude of the parameter $\alpha$. Indeed, if at the bounce
\begin{equation}
\rho_\beta=\rho_{crit}^{vac}(a_B)=-\frac{1}{16\alpha\chi}\sim \rho_{Planck},
\end{equation}
where\footnote{For the sake of clarity here we show explicitly the speed of light $c$, that in the rest of the work we set to unity.} $\rho_{Planck}=c^7/\hbar G^2$, then it follows that $\alpha$ can be estimated by
\begin{equation}
|\alpha|\sim\frac{\hbar G}{128\pi c^3}.
\label{alpha estimate}
\end{equation}
\\We note that the branch $\mathcal{D}_2$ is marked by another peculiar point, namely $a=a_R=(-24\alpha C_0^2)^{1/6}$ where the function $H^2$ diverges, and we have a vanishing Hubble radius for a finite scale factor \cite{Odintsov:2018uaw,Astashenok:2012tv}. We expect that this type of singularity could be regularized taking into account the gravitational particle creation \cite{Montani:2001fp,Dimopoulos:2018kgl,Ford:1986sy}, related to the presence of a cosmological horizon, or the non-equilibrium nature of the involved thermodynamic processes \cite{Disconzi:2014oda,Venanzi:2016pds,Israel:1976tn}, like bulk viscosity effects \cite{Barrow:1988yc,Belinskii1,Belinskii2}. In particular, particle creation can be described by means of additional terms in the Friedman equation, able to stabilize the singular behaviour of the Hubble parameter \cite{Contreras:2018two}. Therefore, we hypothesize that the Universe might evolve smoothly through the critical point $a_R$, reaching asymptotically the General Relativity regime \eqref{R2_GR}, where we also require that the Immirzi field relaxes on a constant value. That can be easily checked combining \eqref{phi_a_R2} and \eqref{eq_beta_FRW}, namely
\begin{equation}
\lim_{a\rightarrow +\infty}\dot{\beta}(a)=\lim_{a\rightarrow +\infty}\frac{a^3(1-2\alpha\chi(3w-1)\rho)}{a^6+6\alpha C_0^2}C_0^2=0,
\end{equation}
which, providing $w\ge-2$, holds irrespective of the specific energy density content $\rho$.
\subsection{The cosmological constant case}\label{cosmologicalconstant}
For $w=-1$ the energy density does not depend on the scale factor and we can formally set $\rho=\Lambda/\chi$, where $\Lambda$ is a cosmological constant term. Then, relation \eqref{phi_a_R2} reads as
\begin{equation}
\phi(a)=\frac{a^6 (1+8\alpha\Lambda)}{a^6+6\alpha C_0^2},
\end{equation}
and \eqref{modified friedman general} takes the form
\begin{equation}
H^2=\frac{(a^6+6\alpha C_0^2)(4\Lambda a^{12}+3C_0^2 a^6+36\alpha C_0^4)}{12a^6(a^6+24 \alpha C_0^2)^2}.
\end{equation}
By close analogy with \eqref{R2_GR}, for $\alpha>0$ it can be simply recast as
\begin{equation}
H^2=\frac{\Lambda_\beta(a)}{3}+\frac{\chi_\beta(a)}{3}\rho_\beta,
\label{modified friedman lambda}
\end{equation}
with the effective cosmological constant $\Lambda_\beta$ given by
\begin{equation}
\Lambda_\beta(a)\equiv \frac{a^6(a^6+6\alpha C_0^2)}{(a^6+24\alpha C_0^2)^2}\;\Lambda,
\label{G eff lambda}
\end{equation}
and for $a\rightarrow +\infty$, the dynamical term $\Lambda_\beta$ asymptotically reaches the constant value $\Lambda$ and the standard de Sitter phase is recovered. On the other hand, near the singularity the $\Lambda$ term is negligible, \textit{i.e.} $\Lambda_\beta\rightarrow 0$ and the Immirzi field is the leading contribution to the dynamics.
\\ If instead $\alpha<0$, the requirement of having a positive Hubble parameter compels once again the scale factor into specific regions of $\mathbb{R}^+$. Specifically, assuming the value of $\Lambda$ fixed, as for instance by current data \cite{Aghanim:2018eyx}, it is possible to distinguish two separate cases, labelled by the size of $\alpha$ with respect to $\Lambda$, \textit{i.e.}:
\begin{equation}
\begin{split}
8\Lambda\alpha<-1&\Rightarrow
\begin{cases}
&\mathcal{D}_1^\Lambda\;a \in [0,a_\Lambda],\\
&\mathcal{D}_2^\Lambda\;a \in [(-6\alpha C_0^2)^{1/6},+\infty);
\end{cases}\\
-1<8\Lambda\alpha<0&\Rightarrow
\begin{cases}
&\mathcal{D}_3^\Lambda\;a \in [0,(-6\alpha C_0^2)^{1/6}],\\
&\mathcal{D}_4^\Lambda\;a \in [a_\Lambda,+\infty);
\end{cases}
\end{split}
\label{constant domains}
\end{equation}
where $a_\Lambda=\leri{-\frac{3C_0^2}{8\Lambda}\leri{1-\sqrt{1-64\alpha\Lambda}}}^{1/6}$.
\\Analogously to the vacuum case, the domains $\mathcal{D}_{1,3}^\Lambda$ always designate closed Universes, which do not admit General Relativity as limit, and they can be overlooked. Instead, branches $\mathcal{D}_{2,4}^\Lambda$ describe bouncing cosmologies, with the Big Bounce point critically depending on the value of $\alpha$. Especially, when $-1<8\Lambda\alpha<0$ holds, the bounce takes place for values corresponding to the turning point $a_T$ of the vacuum case (see \eqref{vacuum domains}), while if $8\Lambda\alpha<-1$ the Big Bounce point is determined by $a_\Lambda$ and also the cosmological constant term is involved in fixing its value. Moreover, it is easy to see that in both cases the bounce occurs for scale factor values lower than in vacuum, being $a_\Lambda<a_B=(-12\alpha C_0^2)^{1/6}$ always satisfied for $\alpha<0$. 
However, if we assume $\alpha$ fixed by \eqref{alpha estimate} and $\Lambda$ reproducing the current dark energy phase ($\Lambda\sim 10^{-18} l_P^{-2}$), then we see that the condition $8\alpha\Lambda<-1$ cannot be satisfied, and $\mathcal{D}_4^{\Lambda}$ is the only valid branch.
\\ Finally, the critical point $a_R=(-24\alpha C_0^2)^{1/6}$, where $H$ diverges, is not removed since for negative values of $\alpha$ it is always contained in the $\mathcal{D}_{2,4}^\Lambda$ domains.

\subsection{The radiation case}\label{radiationcase}
When $w=1/3$ the trace of the stress-energy tensor vanishes and relation \eqref{phi a R2 vacuum} is unaltered, whereas the Friedman equation \eqref{friedman vacuum R2} is slightly modified and reads as:
\begin{equation}
H^2=\frac{C_0^2}{4a^6}\frac{(a^6+6\alpha C_0^2)(a^6+12\alpha C_0^2+\frac{4\chi\mu_R^2}{3C_0^2}\frac{(a^6+6\alpha C_0^2)^2}{a^4})}{(a^6+24 \alpha C_0^2)^2}.
\end{equation}
Following \eqref{modified friedman lambda}, when $\alpha>0$ it can be  rewritten as
\begin{equation}
H^2=\frac{\chi_R(a)}{3}\rho_R+\frac{\chi_\beta (a)}{3}\rho_\beta,
\end{equation}
where with analogy with \eqref{G eff lambda} we defined the effective gravitational coupling
\begin{equation}
\chi_R(a)=\frac{(a^6+6\alpha C_0^2)^3}{a^6(a^6+24\alpha C_0^2)^2}\;\chi.
\label{G eff rad}
\end{equation}
In particular, by virtue of \eqref{G eff imm}-\eqref{G eff rad}, we see that near the singularity the Immirzi energy density is negligible and the Friedman equation behaves like $H^2\sim a^{-10}$, corresponding to an effective superluminal index $w=7/3$. We observe that such results are quite common in ekpyrotic theories (see \cite{Lehners:2008vx} and references therein for an introduction), where it is in general requested $w\gg 1$ in order to solve the fine tuning issues of standard cosmological model.
\\ When $\alpha<0$, it can be demonstrated with bit long calculations that the effect of the radiation energy density is twofold: It endows the Hubble function of an additional zero $a_{B_1}$ with respect to the vacuum case and displaces the critical point $a_B=(-12\alpha C_0^2)^{1/6}$ in a new root $a_{B_2}$. Even if such two new zeros cannot be analytically evaluated, they may be still algebraically estimated by
\begin{equation}
\begin{split}
&a_{B_1}\in\leri{0;(-6\alpha C_0^2)^{1/6}}\\
&a_{B_2}\in\leri{(-6\alpha C_0^2)^{1/6};(-12\alpha C_0^2)^{1/6}}.
\end{split}
\end{equation}
Accordingly, the regions where relation $H^2\ge 0$ is valid are changed into the following new domains:
\begin{equation}
\begin{split}
&\mathcal{D}_1^R:\;a \in [a_{B_1},(-6\alpha C_0^2)^{1/6}],\\
&\mathcal{D}_2^R:\;a \in [a_{B_2},+\infty),
\label{radiation domains}
\end{split}
\end{equation}
and we see that the unphysical branch $\mathcal{D}_1^R$ is now turned in a cyclic Universe equipped with a proper bounce point. Concerning instead $\mathcal{D}_2^R$, we note that the Big Bounce is shifted to lower values, as for the cosmological constant case, whereas the point of divergence $a_R$ is unaffected.

\section{Concluding remarks}
\label{sec6}
The analysis above provided the cosmological implementation of the Palatini $f(R)$ model discussed in \cite{Bombacigno:2018tbo}, where a Nieh-Yan term was included in the presence of an Immirzi field. The peculiarity of that approach was the possibility to completely solve torsion in terms of the Immirzi and gravitational d.o.f., so obtaining a non-minimally coupled scalar-tensor theory.
\\ In particular, we considered a specific class of $f(R)$ theories, mimicking the well-established Starobinsky model in metric formalism, by means of a quadratic correction to the Palatini-Hilbert action. In this regard, we clearly distinguished two different cosmological scenarios, depending on the sign of such a correction. Indeed, for $\alpha>0$ the analysis outlined a slightly modified Friedman dynamics, approaching in the asymptotic limit the standard description of general relativity and characterized by effective gravitational couplings, according the type of energy density considered. For $\alpha<0$ instead, we pointed out the existence of radically different solutions, consisting in closed and bouncing Universes, respectively. Especially, the former were obtained even for vanishing spatial curvature, but in general they turned out to be still singular, and they were ruled out because of the absence of the general relativity limit. Concerning the latter, we were able to identify in the non minimal coupling of the Immirzi field with the gravitational d.o.f. the cause of the classical removal of the initial singularity. In this respect, it is worth stressing that when a radiation energy density was taken into account, the combined effect of the Immirzi and radiation field was to introduce a further bouncing point in the closed solution, resulting in a cyclic model. We mention that this kind of solutions, even if \textit{a priori} disregarded, they could represent a Planckian state of the Universe, from which the bouncing branch could originate as the result of a quantum tunneling phenomenon.
\\ Moreover, we pointed out that in general the reliable classical solutions are always endowed with critical points associated to little rip dynamics, where the Friedman equation diverges for specific values of the scale factor. Of course, they must be regularized by reducing to a finite value the expansion rate, as effect of matter creation, as well as non equilibrium thermodynamics implications, mainly associated to bulk viscosity effects.
\\ Eventually, we shown that the Immirzi field can asymptotically relaxed to a constant by the expansion of the Universe, newly recovering the LQG vision of an Immirzi parameter.


\begin{thebibliography}{100}
\bibitem{CQG}
F.~Cianfrani, O.M.~Lecian, M.~Lulli, G.~Montani, \textit{Canonical quantum gravity} (World Scientific Pub Co Inc., Singapore, 2014)

\bibitem{Nojiri:2010wj} 
  S.~Nojiri and S.~D.~Odintsov,
  Phys.\ Rept.\  {\bf 505}, 59 (2011)


\bibitem{Sotiriou:2008rp}
  T.~P.~Sotiriou and V.~Faraoni,
  Rev.\ Mod.\ Phys.\  {\bf 82} (2010) 451

  
\bibitem{Nojiri:2017ncd}
  S.~Nojiri, S.~D.~Odintsov and V.~K.~Oikonomou,
  Phys.\ Rept.\  {\bf 692} (2017) 1


\bibitem{ST}
Y.~Fujii and K.~Maeda,  \textit{The Scalar-tensor theory of Gravitation} (Cambridge University Press, Cambridge, 2009)

\bibitem{Pal}
A.~Palatini, Rend. Circ. Mat. Palermo {\bf 43}, 203 (1919) [English translation by R. Hojman and C. Mukku, in \textit{Cosmology and Gravitation}, edited by P.G. Bergmann and V. De Sabbata (Plenum Press, New York, 1980)]

\bibitem{Olmo:2011uz} 
  G.~J.~Olmo,
  Int.\ J.\ Mod.\ Phys.\ D {\bf 20}, 413 (2011)

\bibitem{Cartan1}
\'{E}.~Cartan. Sur une g\'{e}n\'{e}ralisation de la notion de courbure de Riemann et les espaces \`{a} torsion, \textit{Cosmology and Gravitation: Spin, Torsion, Rotation, and Supergravity}, Nato Science Series B, edited by P.G. Bergmann and V. De Sabbata (Springer, New York, 1980), p. 489.

\bibitem{Cartan2}
\'{E}.~Cartan. Ann. Sci. \'{E}cole Norm. Sup. {\bf 40}, 325 (1923); Ann. Sci. Éc. Norm. Sup. {\bf 41}, 1 (1924); Ann. Sci. \'{E}cole Norm. Sup. {\bf 42}, 17–88 (1925).

\bibitem{Hehl:1976kj}
  F.~W.~Hehl, P.~Von Der Heyde, G.~D.~Kerlick and J.~M.~Nester,
  Rev.\ Mod.\ Phys.\  {\bf 48}, 393, (1976) .

\bibitem{Shapiro:2001rz}
  I.~L.~Shapiro,
  Phys.\ Rept.\  {\bf 357} (2002) 113

\bibitem{Hehl:2013qga} 
  F.~W.~Hehl, Y.~N.~Obukhov and D.~Puetzfeld,
  Phys.\ Lett.\ A {\bf 377}, 1775 (2013)
 
\bibitem{Bombacigno:2018tbo} 
  F.~Bombacigno and G.~Montani,
  Phys.\ Rev.\ D {\bf 97}, 124066 (2019)

\bibitem{Ashtekar1}
A.~Ashtekar, Phys.\ Rev.\ Lett {\bf 57}  2244 (1986)

\bibitem{Barbero1}
J.F.~Barbero G., Phys.\ Rev.\ Lett {\bf 51} 5507 (1995)

\bibitem{Holst:1995pc}
  S.~Holst,
  Phys.\ Rev.\ D {\bf 53} 5966 (1996)

\bibitem{Immirzi:1996di} 
  G.~Immirzi,
  Class.\ Quant.\ Grav.\  {\bf 14}, L177 (1997) 


\bibitem{Thiemann:2007zz} 
 T.~Thiemann, \textit{Modern canonical quantum general relativity} (Cambridge University Press, Cambridge, 2007)

  
\bibitem{Rovelli}
C.~Rovelli, \textit{Quantum Gravity} (Cambridge University Press, Cambridge, 2004)


\bibitem{NY1}
H.T. Nieh and M.L. Yan, J. Math. Phys. 23, 373-374 (1982);

\bibitem{NY2}
H.T. Nieh, Int. J. Mod. Phys. A22, 5237-5244 (2007).

\bibitem{Mercuri:2007ki} 
  S.~Mercuri,
  Phys.\ Rev.\ D {\bf 77}, 024036 (2008)

\bibitem{Rovelli:1997na} 
  C.~Rovelli and T.~Thiemann,
  Phys.\ Rev.\ D {\bf 57}, 1009 (1998)

\bibitem{Calcagni:2009xz} 
  G.~Calcagni and S.~Mercuri,
  Phys.\ Rev.\ D {\bf 79}, 084004 (2009)
  
\bibitem{TorresGomez:2008fj} 
  A.~Torres-Gomez and K.~Krasnov,
  Phys.\ Rev.\ D {\bf 79}, 104014 (2009)
 
\bibitem{Cianfrani:2009sz} 
  F.~Cianfrani and G.~Montani,
  Phys.\ Rev.\ D {\bf 80}, 084040 (2009)

\bibitem{Bombacigno:2016siz}
  F.~Bombacigno, F.~Cianfrani and G.~Montani,
  Phys.\ Rev.\ D {\bf 94} (2016) no.6,  064021

\bibitem{Bombacigno:2018tih} 
  F.~Bombacigno and G.~Montani,
  Phys.\ Rev.\ D {\bf 99}, 064016 (2019)
  
\bibitem{Antoniadis:2018ywb} 
  I.~Antoniadis, A.~Karam, A.~Lykkas and K.~Tamvakis,
  JCAP {\bf 1811}, no. 11, 028 (2018)

\bibitem{Tenkanen:2019jiq} 
  T.~Tenkanen,
  arXiv:1901.01794 [astro-ph.CO].
  
\bibitem{Edery:2019txq} 
  A.~Edery and Y.~Nakayama,
  arXiv:1902.07876 [hep-th].

\bibitem{Ghosh:2004wq} 
  A.~Ghosh and P.~Mitra,
  Phys.\ Lett.\ B {\bf 616}, 114 (2005)

\bibitem{Ghosh:2011fc} 
  A.~Ghosh and A.~Perez,
  Phys.\ Rev.\ Lett.\  {\bf 107}, 241301 (2011)
  Erratum: [Phys.\ Rev.\ Lett.\  {\bf 108}, 169901 (2012)]

\bibitem{Starobinsky}
A.A.~ Starobinsky, Phys.\ Lett.\ B, {\bf 91}, 99 (1980)

\bibitem{Hawking:2000bb} 
  S.~W.~Hawking, T.~Hertog and H.~S.~Reall,
  Phys.\ Rev.\ D {\bf 63}, 083504 (2001)

\bibitem{Kehagias:2013mya} 
  A.~Kehagias, A.~Moradinezhad Dizgah and A.~Riotto,
  Phys.\ Rev.\ D {\bf 89}, no. 4, 043527 (2014)


\bibitem{Whitt:1984pd} 
  B.~Whitt,
  Phys.\ Lett.\  {\bf 145B}, 176 (1984).

\bibitem{Sotiriou:2006sf} 
  T.~P.~Sotiriou,
  Phys.\ Lett.\ B {\bf 645}, 389 (2007)


\bibitem{Olmo:2008nf} 
  G.~J.~Olmo and P.~Singh,
  JCAP {\bf 0901}, 030 (2009)
  
\bibitem{Barragan:2009sq} 
  C.~Barragan, G.~J.~Olmo and H.~Sanchis-Alepuz,
  Phys.\ Rev.\ D {\bf 80}, 024016 (2009)
  
\bibitem{Koivisto:2010jj} 
  T.~S.~Koivisto,
  Phys.\ Rev.\ D {\bf 82}, 044022 (2010)  
  

\bibitem{Odintsov:2018uaw} 
  S.~D.~Odintsov and V.~K.~Oikonomou,
  Phys.\ Rev.\ D {\bf 98}, no. 2, 024013 (2018)

\bibitem{Astashenok:2012tv} 
  A.~V.~Astashenok, S.~Nojiri, S.~D.~Odintsov and A.~V.~Yurov,
  Phys.\ Lett.\ B {\bf 709}, 396 (2012)

\bibitem{Montani:2001fp} 
  G.~Montani,
  Class.\ Quant.\ Grav.\  {\bf 18}, 193 (2001)

\bibitem{Dimopoulos:2018kgl} 
  K.~Dimopoulos,
  Phys.\ Lett.\ B {\bf 785}, 132 (2018)

\bibitem{Ford:1986sy} 
  L.~H.~Ford,
  Phys.\ Rev.\ D {\bf 35}, 2955 (1987).

\bibitem{Disconzi:2014oda} 
  M.~M.~Disconzi, T.~W.~Kephart and R.~J.~Scherrer,
  Phys.\ Rev.\ D {\bf 91}, no. 4, 043532 (2015)
 

\bibitem{Venanzi:2016pds} 
  G.~Montani and M.~Venanzi,
  Eur.\ Phys.\ J.\ C {\bf 77}, no. 7, 486 (2017)

\bibitem{Israel:1976tn} 
  W.~Israel,
  Annals Phys.\  {\bf 100}, 310 (1976).



\bibitem{Barrow:1988yc} 
  J.~D.~Barrow,
  Nucl.\ Phys.\ B {\bf 310}, 743 (1988).

\bibitem{Belinskii1}
V.A.~Belinskii and I.M.`Khalatnikov, 
 Sov.\ J.\ Exp.\ Theor. Phys. {\bf 42}, 205 (1975)
 
 
\bibitem{Belinskii2}
V.A.~Belinskii, E.S.~Nikomarov and I.M.~Khalatnikov, 
Sov.\ J.\ Exp.\ Theor.\ Phys. {\bf 50} (2), 21 (1979)

\bibitem{Contreras:2018two} 
  F.~Contreras, N.~Cruz, E.~Elizalde, E.~González and S.~Odintsov,
  arXiv:1808.06546 [gr-qc].

\bibitem{Aghanim:2018eyx} 
  N.~Aghanim {\it et al.} [Planck Collaboration],
  arXiv:1807.06209 [astro-ph.CO].
  
\bibitem{Lehners:2008vx} 
  J.~L.~Lehners,
  Phys.\ Rept.\  {\bf 465}, 223 (2008)


\end{thebibliography}
\end{document}